\documentclass[reprint,aps,superscriptaddress,nofootinbib,groupedaddress,preprintnumbers]{revtex4-1}

\usepackage{amsmath,amsthm,amsfonts}
\usepackage{graphicx}
\usepackage{physics}
\usepackage{hyperref}
\hypersetup{
	colorlinks   = true, 
	urlcolor     = blue, 
	linkcolor    = blue, 
	citecolor    = red   
}

\DeclareMathSymbol{\mlq}{\mathord}{operators}{``}
\DeclareMathSymbol{\mrq}{\mathord}{operators}{`'}

\graphicspath{{figs/}}

\newcommand{\nc}{\newcommand}

\nc{\calR}{{\cal{R}}}
\nc{\calP}{{\cal{P}}}
\nc{\cN}{ {\cal{N}} }

\nc{\Mpt}{M_{_{\rm Pl}}^2}

\usepackage{tikz}
\usetikzlibrary{arrows,shapes}
\usetikzlibrary{trees}
\usetikzlibrary{matrix,arrows} 				
\usetikzlibrary{positioning}				
\usetikzlibrary{calc,through}				
\usetikzlibrary{decorations.pathreplacing}  
\usepackage{pgffor}							

\usetikzlibrary{decorations.pathmorphing}	
\usetikzlibrary{decorations.markings}
\tikzset{
    vector/.style={decorate, decoration={snake}, draw},
	provector/.style={decorate, decoration={snake,amplitude=2.5pt}, draw},
	antivector/.style={decorate, decoration={snake,amplitude=-2.5pt}, draw},
    fermion/.style={draw=black, postaction={decorate},
        decoration={markings,mark=at position .55 with {\arrow[draw=black]{>}}}},
    fermionbar/.style={draw=black, postaction={decorate},
        decoration={markings,mark=at position .55 with {\arrow[draw=black]{<}}}},
    fermionnoarrow/.style={draw=black},
    gluon/.style={decorate, draw=black,
        decoration={coil,amplitude=4pt, segment length=5pt}},
    scalar/.style={dashed,draw=black, postaction={decorate},
        decoration={markings,mark=at position .55 with {\arrow[draw=black]{>}}}},
    scalarbar/.style={dashed,draw=black, postaction={decorate},
        decoration={markings,mark=at position .55 with {\arrow[draw=black]{<}}}},
    scalarnoarrow/.style={dashed,draw=black},
    electron/.style={draw=black, postaction={decorate},
        decoration={markings,mark=at position .55 with {\arrow[draw=black]{>}}}},
	bigvector/.style={decorate, decoration={snake,amplitude=4pt}, draw},
}

\tikzstyle{block} = [draw, rectangle, 
    minimum height=3em, minimum width=6em]

\begin{document}


\title{Probing single-field inflation: predictions, constraints, and theoretical viewpoints}
\author{Fereshteh Felegary,$^{1}$ Thammarong Eadkhong,$^{1}$, Farruh Atamurotov,$^{2,3}$ and Phongpichit Channuie$^{1,4}$}
\email{fereshteh.felegary@gmail.com,phongpichit.ch@mail.wu.ac.th}
 
 \affiliation{$^{1}$School of Science, Walailak University, Nakhon Si Thammarat, 80160, Thailand}

 \affiliation{$^{2}$University of Tashkent for Applied Sciences, Str. Gavhar 1, Tashkent 100149, Uzbekistan}
\affiliation{$^{3}$Urgench State University, Kh. Alimdjan str. 14, Urgench 220100, Uzbekistan}

\affiliation{$^{4}$College of Graduate Studies, Walailak University, Nakhon Si Thammarat, 80160, Thailand}

\begin{abstract}
This work investigates a single-field inflationary model, a specific class of the K-essence models where a coupling term exists between canonical Lagrangian and the potential. This coupling term has many effects on main inflationary parameters consisting of the power spectral,  the spectral index, the tensor-to-scalar ratio, the Hubble parameter, the equation of state parameter, and the slow-roll parameter. By solving the equations numerically and deriving analytical results, how this modification affects inflationary dynamics can be analyzed. Our results show that the coupling term, $\alpha$, decreases the inflationary parameters, such as the tensor-to-scalar ratio, $r$, and improves the consistency with observational constraints from Planck and BICEP/Keck at the $68 \%$ and $95 \%$ confidence. These findings indicate that the studied model provides a promising alternative to the early universe dynamics while aligning with recent cosmological observations.
\end{abstract}

\maketitle 
\section{Introduction}\label{sec0}
Cosmology, a comprehensive study of the beginning and evolution of the universe, has been included as a branch of Physics science in the  20th century. In the $1920$s, the Hot Big Bang (HBB) was released, and General Relativity (GR) provided its theoretical basis in which the universe has been expanding from a much hot and dense initial state in a way far past \cite{Einstein:1916vd, Lemaitre:1927zz}. Utilizing several observational discoveries, such as the Cosmic Microwave Background (CMB) and the universe's expansion, the HBB was the cosmology standard model \cite{Hubble:1929ig, penzias1965medicion}. The expanded universe gets cold almost $380,000$ years after the HBB, generating the CMB. In the 1980s, it was found that vital primary conditions for successful Hot Big Bang theory in cosmology could be created by a period of fast expansion of the cosmos in the early times that is named cosmic inflation \cite{PhysRevD.23.347, Linde:1981mu, Albrecht:1982wi}. Moreover, inflation responds to the main questions of the cosmos, such as the fine-tuning and horizon problems, and provides the primordial seed of the structure formation of the cosmos \cite{Mukhanov:1981xt}. Because of the rapid expansion during inflation, quantum mechanics fluctuations are stretched to get classical perturbations. At the end of the cosmic inflation, these perturbations become the seed for the temperature anisotropies in the CMB. Hence, we can precisely peruse the primordial cosmic inflation by addressing this issue. 

Inflation describes the physics of the early times, but the nature of the field in charge of inflation is still an important question. Generally, in an inflationary universe, the inflaton field $\phi$ is supposed to drive the accelerated expansion. During the inflation era, this field dominates the energy density of the universe. It is worth noting that inflation is usually brought up in terms of potential energy, which is a function of the scalar field. Also, the validity of inflationary models has been inquired by observational data, including the CMB, Large-Scale Structure (LSS), and so forth \cite{Morishita:2022bkr}.

In inflationary cosmology, two major quantities play a fundamental role in the early universe as the inflationary parameters: the tensor-to-scalar ratio, which specifies the competitive contribution of primordial Gravitational Waves (GWs) to the CMB anisotropies, and the spectral index, which represents the scale dependence of primordial fluctuations \cite{Kolb:1994ur}.
Recently, the findings from $Planck$ remarkably restrict the limitation of acceptable inflationary parameters in the early era. For instance, the observational constraints from $Planck$ and $BICEP/Keck$ point out that there is a bound  $r<0.036$ with $95\% ~ C.L.$ confidence for the tensor-to-scalar ratio parameter \cite{BICEP:2021xfz} and the observational constraints from $Planck$ point out that there is a bound $n_{s}=0.9649 \pm 0.0042$ with $68\% ~ C.L.$ confidence for the spectral index \cite{Planck:2018jri}. Therefore, the single-field models are among the most extensively studied topics in early universe cosmology, and these models depend on their potential \cite{Wolf:2024lbf, PhysRevD.82.043502, PhysRevD.100.043522, PhysRevD.90.047303, Liddle:2000cg, Komatsu_2011, Senatore:2009gt, Unnikrishnan:2013vga, Unnikrishnan:2012zu, Li_2012, Akarsu:2023nyl}.

Some studies have figured out that several chaotic inflation models with potential $V(\phi) \propto \phi^{n}$ are ruled out by the last findings \cite{Linde:1981mu, Linde:1983gd}. To solve this problem, some inflationary models have been considered, such as the energy-momentum squared gravity (EMSG) model \cite{HosseiniMansoori:2023zop}, $\alpha-$attractor model \cite{Ferrara:2013rsa}, the new relativistic theory of Modified Newtonian Dynamics (MOND) \cite{Bekenstein:2004ne}, and so on.
\\
Our main aim of this work is to consider an inflation single-field model in which some important quantities of the cosmos background and the primordial cosmic perturbations such as the Hubble parameter, $H$, Equation of State parameter (EoS),  $\omega$, the Slow-roll parameters, $\epsilon_{{H}}$ and $\eta_{{H}}$, the Power Spectrum, $\mathcal{P}_{\mathcal{R}}$, the tensor-to-scalar ratio, $r$ and the spectral index, $n_{s}$ align with the recent observational constraints. For this target, we consider a specific class of the K-essence models, which are described by a coupling term between the canonical Lagrangian and the potential \cite{Armendariz_Picon_2001}.
The existence of this coupling term in this model makes adjustments to the slow-roll dynamics, which play a determined role in the inflationary evolution of the cosmos. Particularly, this coupling term modifies the evolution of the slow-roll parameters and affects the evolution of the inflaton field. Therefore,  important inflationary parameters such as the spectral index and the tensor-to-scalar ratio are affected by this modification. As a result, these parameters are essential to characterize the nature of primordial fluctuations and comprehension of the influence of the coupling term. By using the numerical and analytical findings, we predict that this model can generate great values for $n_{s}$ and $r$ align with closely from Planck and BICEP/Keck. This represents that the presence of this coupling term provides a good extension to standard single-field inflationary scenarios, which is in agreement with the last observational constraints in the early universe, and meanwhile, there is a delicate description.

The framework of this paper is formed as follows: 
In section \ref{sec1}, we present our model and derive the background dynamics of its single-field model. In the next section, utilizing the equations of the background dynamics, we obtain the main inflationary parameters for our model. In section \ref{sec3}, we investigate the numerical and Analytical study of early universe parameters. In the next section, we inquire about the numerical solutions and analytical predictions for our model. Finally, In the last section, we summarize our findings of the studied model.
\section{The single-field Model and background dynamics}\label{sec1}
We start by defining the action of the single-field model as follows  \cite{Chen:2006nt}:
\begin{equation}\label{action}
\mathcal{S} = \int d^{4}x \sqrt{-g}\Big[\frac{R}{2 m_{pl}^{2}} + P(X , \phi)\Big],
\end{equation}
$m_{pl}$ is the reduced Planck mass, $R$ is the Ricci scalar, and $P(X, \phi)$ is an arbitrary function of the kinetic term and the single-scalar field. 
Note that we have set $m_{pl}=1$ in this work. Also, the kinetic term is defined as  \cite{Chen:2006nt}:
\begin{equation}\label{X}
X = -\frac{1}{2}\partial_{\mu}\phi \partial^{\mu}\phi.
\end{equation}
We also consider the spatially flat Friedmann-Lemaitre-Robertson-Walker (FLRW) space-time in this work:
\begin{equation}
ds^{2}=-dt^{2}+a^{2}(t) \delta^{ij} dx_{i}dx_{j}.
\end{equation}
Here, $a$ and $\phi$ are the scale factor and the scalar-single field, respectively.  These depend only on cosmic time, $t$. Let us consider a specific class of the $\textit{K-essence}$ models which is described by \cite{Armendariz_Picon_2001}:
\begin{equation}\label{P}
P(X , \phi) = \mathcal{F}(\phi)\Big[X - V(\phi)\Big],
\end{equation}
where $V(\phi)$ is a potential function of the scalar field and $\mathcal{F}(\phi)$ is a functional of the potential which is defined by
\begin{equation}\label{f}
\mathcal{F}(\phi) = 1-\frac{\alpha}{m_{pl}^{2}} V(\phi),
\end{equation}
where $\alpha$ is a dimensionless constant. It is necessary to mention that we have set  $\alpha>0$ and the upper bound on the potential as $V<1/\alpha$ at the level of cosmological perturbation. The reason for choosing these conditions is to remedy the ghost and the gradient instabilities. In addition, for  $\alpha=0$, this model reverts to the standard canonical single-field model.

The energy-momentum tensor, $T_{\mu\nu}$, can be derived by varying the action, Eq. (\ref{action}), with respect to the metric, $g^{\mu\nu}$. Therefore, $T_{\mu\nu}$ is described as  
\cite{Gong:2016qmq}:
\begin{equation}\label{energymomentum}
T_{\mu\nu}\equiv -\frac{2}{\sqrt{-g}} \frac{\delta \mathcal{S}}{\delta g^{\mu \nu}}
=P g_{\mu\nu} + P_{,X} \partial_{\mu}\phi \partial_{\nu}\phi,
\end{equation}
which in $P_{,X}$ denotes the partial derivative of $P$ with respect to $X$.
In the flat space-time, by substituting Eq. (\ref{P}) into Eq. (\ref{energymomentum}), one obtains
\begin{equation}\label{energymomentummulti1}
T_{\mu\nu}=  \mathcal{F}(\phi)\Big[X - V\Big] g_{\mu\nu} +  \mathcal{F}(\phi) \partial_{\mu}\phi \partial_{\nu}\phi.
\end{equation}
In a perfect fluid, the scalar field is homogeneous, and using Eq. (\ref{energymomentummulti1}), the energy density of the scalar field is defined by  \cite{Chen:2006nt}:
\begin{equation}\label{rho}
\rho(X , \phi) = 2XP_{,X} - P = \mathcal{F}(\phi)\Big[X + V(\phi)\Big].
\end{equation}
In addition, the background evolution is governed by a set of cosmological equations \cite{Chen:2006nt}
\begin{equation}\label{H2}
H^{2}=\frac{1}{3}\Big[2XP_{,X} - P \Big] = \frac{1}{3}\mathcal{F}(\phi)\Big[X + V(\phi)\Big],
\end{equation}
\begin{equation}\label{dotH}
\dot{H} = -X P_{,X} = -X \mathcal{F}(\phi),
\end{equation}
where $H=\dot{a}/a$ is called the Hubble parameter and the dot symbol denotes the derivative with respect to the cosmic time. 
In the flat space-time,  by varying the action, Eq. (\ref{action}), with respect to $\phi$  and using Eqs. (\ref{P}) and (\ref{f}), the equation of motion can be derived by
\begin{equation}\label{ddotphi}
\ddot{\phi} + \Big[3H + \frac{\dot{\phi} \mathcal{F}_{,\phi}}{2\mathcal{ F}} \Big] \dot{\phi} + \frac{\mathcal{F}_{,\phi}}{\mathcal{F}} V + V_{,\phi} =0.
\end{equation}
Here, the subscript $\mlq\mlq~~_{,\phi} ~~\mrq\mrq$ denotes $d/d\phi$. As expected, for $\alpha=0$,  the equation of motion reduces to the standard canonical single-field model.

The square of an effective sound speed is defined  \cite{Chen:2006nt}:
\begin{eqnarray}\label{soundspead}
c_{s}^{2} = \frac{P_{,X}}{P_{,X} + 2 X P_{,XX}},
\end{eqnarray}
For our model, by substituting Eq. (\ref{P}) into Eq. (\ref{soundspead}), the square of the effective sound speed equals one i.e. $c_{s}^{2}=1$. 
\section{Main Inflationary Parameters}\label{sec2}
In this section, we would like to study a period of inflation in the early universe by utilizing the background dynamics. To begin with, we represent one of the main inflationary parameters: The number of e-folds represents how much inflation happened to the end of the inflation and is introduced by the following   \cite{Chen:2006nt}:
\begin{equation}\label{Nmulti}
N = \int_{t_{i}}^{t_{f}} H dt=  \int_{\phi_{i}}^{\phi_{f}}\frac{H}{\dot{\phi}}d\phi.
\end{equation}
Let us consider the slow-roll approximation i.e., $\dot{\phi} \ll V $ and $\ddot{\phi} \ll H \dot{\phi}$, then, using Eq. (\ref{f}), Eqs. (\ref{H2}), (\ref{dotH}) and (\ref{ddotphi}) reduce to
\begin{equation}\label{H2multi}
H^{2}\simeq \frac{1}{3}\mathcal{F} V,
\end{equation}
\begin{equation}\label{dotHmulti}
\dot{H} = -X \mathcal{F} ,
\end{equation}
\begin{equation}\label{ddotphimulti}
3H \dot{\phi} \simeq  - \frac{(2\mathcal{F}-1)}{\mathcal{F}}V_{,\phi}.
\end{equation}
By substituting Eqs. (\ref{H2multi}) and (\ref{ddotphimulti}) into Eq. (\ref{Nmulti}), we obtain
\begin{eqnarray}\label{Nnew}
N  \simeq \int_{\phi_{i}}^{\phi_{f}} \Big( \frac{\mathcal{F}^{2}}{(2\mathcal{F}-1)}\Big)\Big(\frac{V}{V_{,\phi}} \Big)d\phi.
\end{eqnarray}
It should be pointed out that there are several essential parameters in the duration of the inflation, such as the slow-roll parameters $(\varepsilon_{H}, \eta_{H})$ and the equation of state parameter $(\omega)$. In this period, the slow-roll parameters must be much smaller than one. This condition is true for 50-60 e-folds to resolve several cosmological problems, such as the horizon and the flatness problems.
Using Eqs. (\ref{H2multi}), (\ref{dotHmulti}) and (\ref{ddotphimulti}), the Hubble slow-roll parameter $\varepsilon_{H}$ finds
\begin{equation}\label{epsilonH}
\varepsilon_{H} = -\dot{H}/H^{2} \simeq \frac{1}{2} \Big(\frac{V_{,\phi}}{V}\Big)^{2}\Big[ \frac{(2\mathcal{F}-1)^{2}}{\mathcal{F}^{3}}\Big].
\end{equation}
By taking the derivative of the Eq. (\ref{epsilonH}) with respect to the time and using Eq. (\ref{epsilonH}) , another Hubble slow-roll parameter $\eta_{H}$ obtains:
\begin{eqnarray}\label{etaH}
\eta_{H} = \frac{\dot{\varepsilon}}{H\varepsilon_{H}}
\simeq   \Big(\frac{V_{,\phi}}{V}\Big)^{2} \Big( \frac{3\mathcal{F} + 6\mathcal{F}^{3} - 7\mathcal{F}^{2}}{\mathcal{F}^{4}} \Big) 
~~~~~~~~~\nonumber\\
- \Big( \frac{2(2\mathcal{F}-1)}{\mathcal{F}^{2}}\Big) \Big(\frac{V_{,\phi\phi}}{V}\Big), ~~~~~~~~~~~~~~~
\end{eqnarray}
where the subscript $\mlq\mlq~~_{,\phi\phi} ~~\mrq\mrq$ denotes $d^{2}/d\phi^{2}$. The equation of state parameter plays a critical role in qualifying the dynamics of the inflationary universe. This parameter is the pressure-to-energy density ratio \cite{Ilic:2010zp}:
\begin{equation}\label{omega}
\omega = \frac{P(X , \phi)}{\rho(X , \phi)}.
\end{equation}
Using Eqs. (\ref{P}), (\ref{rho}),  (\ref{dotHmulti}) and (\ref{epsilonH}) and substituting  into Eq. (\ref{omega}), one finds \cite{Ilic:2010zp}:
\begin{equation}\label{omega1}
\omega = \frac{2}{3} \varepsilon_{H} -1.
\end{equation}
To derive the equations necessary for inflationary parameters, we need to bring up the power-law potentials. These potentials have attracted important interest due to their simplicity and adjustability in explaining early universe dynamics in inflationary cosmology. They, which are specified by a term proportional to the power of the single field, $V(\phi) \sim  \phi^{n}$, suggest a framework to model in the early universe \cite{Linde:1981mu, Linde:1983gd}. Therefore, we consider:
\begin{equation}\label{VV}
V(\phi) =\frac{1}{2} m^{2}\phi^{\frac{2}{5}},
\end{equation}
\begin{equation}\label{VVV}
 V(\phi) =\frac{1}{2} M^{2}\phi,
\end{equation}
where $m$ and $M$ are the constant parameters.
\section{Early Universe Parameters}\label{sec3}
In this section, we aim to investigate the effect of the coupling term $\alpha$ on inflationary parameters, including the tensor-to-scalar ratio and the spectral index in the early universe. Let us utilize the equations of cosmological perturbations in the comoving gauge  \cite{Chen:2006nt, Garriga:1999vw}.
\subsection{Numerical study of early universe parameters}
 We get started by considering the first-order scalar perturbations
to the flat FLRW metric, which is given by  \cite{Chen:2006nt}:
\begin{eqnarray}
\delta g_{00} = 2A ~,~\delta g_{0i} = 2 a \partial_{i}B~,~\delta g_{ij} = a^{2} ( e^{2\mathcal{G}}\delta_{ij} + h_{ij})
\end{eqnarray}
where $A$, $B$, $\mathcal{G}$, and $h_{ij}$  are scalar and the tensor perturbations, respectively.
It should be mentioned that to acquire the first-order equation of motion for the cosmological perturbations, $\delta \phi$, we require to expand the action  Eq. (\ref{action}) up to second-order \cite{Chen:2006nt}
\begin{eqnarray}\label{actionorder2}
\mathcal{S}^{(2)} =\frac{1}{2} \int dt~ d^{3}x~ a^{3}\bigg[\varepsilon_{H} \Big( \dot{\mathcal{G}}^{2} - \frac{(\partial \mathcal{G})^{2}}{a^{2}}\Big)~~~~~~~ \nonumber\\
+ \frac{1}{4} \Big((\dot{h_{ij}})^{2} -\frac{(\partial h_{ij})^{2}}{a^{2}}\Big)
\bigg].
\end{eqnarray}
From Eq. (\ref{actionorder2}), for the perturbations to remain stable and be free of ghost and gradient instabilities, $ \mathcal{F}>0$ must be satisfied. Therefore, this function plays a critical role in maintaining the stability of the inflationary framework.

\begin{figure}
\centering
	\includegraphics[width=0.44\textwidth]{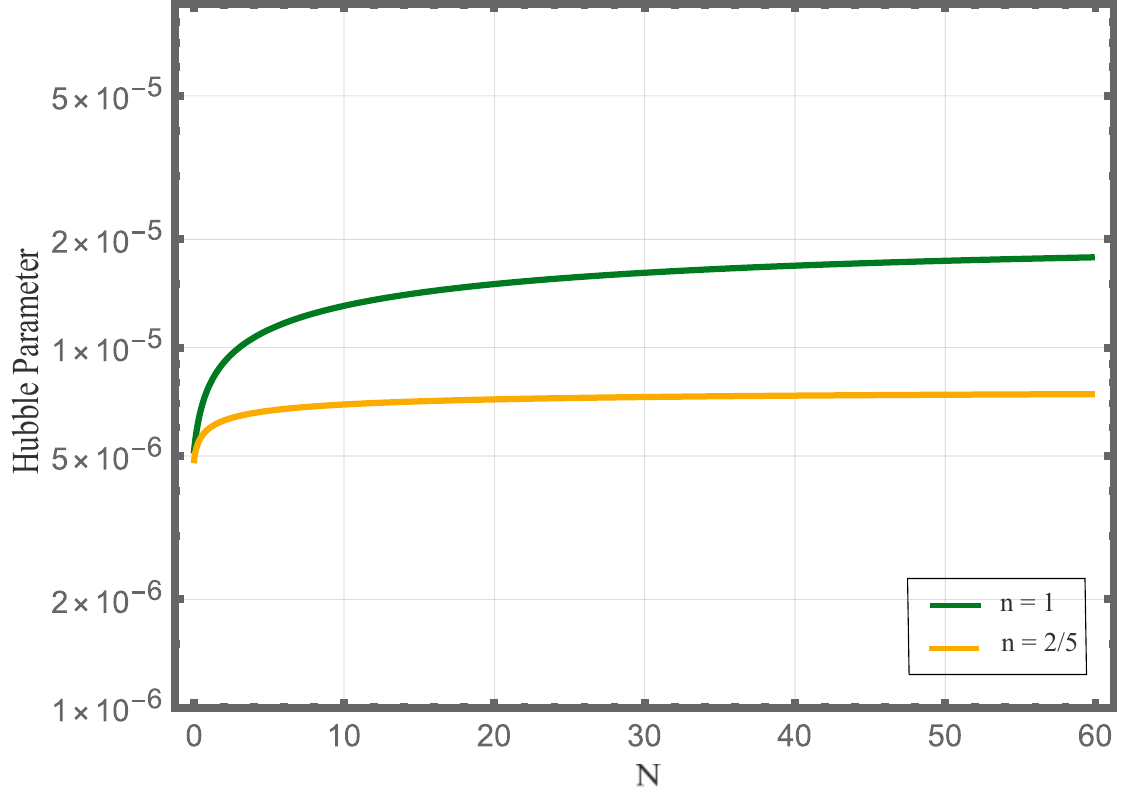}
	\caption{ The evolution of the Hubble parameter as a function of the number of e-folds for $V(\phi) \sim  \phi^{n}$, $N= 60$, $\beta = 0.212$ and $\tilde{\beta} = 0.031$. }	
	\label{fighubble}
\end{figure}

By solving Eqs. (\ref{dotH}) and (\ref{ddotphi}), we can plot the evolution of the Hubble parameter as a function of the number of e-folds for our model, which has been shown in Figure (\ref{fighubble}). In the duration of inflation, the Hubble parameter plays an important role in specifying the expansion rate of the cosmos. This parameter remains almost constant. This is because the potential energy of the inflation field dominates its kinetic energy. 

The power spectrum is one of the key physical quantities during the inflation era. This quantity plays a critical role in describing the Cosmic Microwave Background (CMB) radiation and the distribution of primordial cosmological perturbations. The power spectrum of curvature perturbations in the slow-roll approximation is defined as \cite{Chen:2006nt}:
\begin{equation}\label{scalarcurvsingle}
\mathcal{P}_{\mathcal{R}} \simeq \frac{H^{2}}{8\pi^{2}\varepsilon_{H}}
\end{equation}
The power spectrum of tensor perturbations also is defined as \cite{Mukhanov:2005sc}
\begin{equation}\label{tensorcurvsingle}
\mathcal{P}_{\tau} = \frac{2}{\pi^{2}}H^{2}.
\end{equation}
\begin{figure}
\centering
	\includegraphics[width=0.44\textwidth]{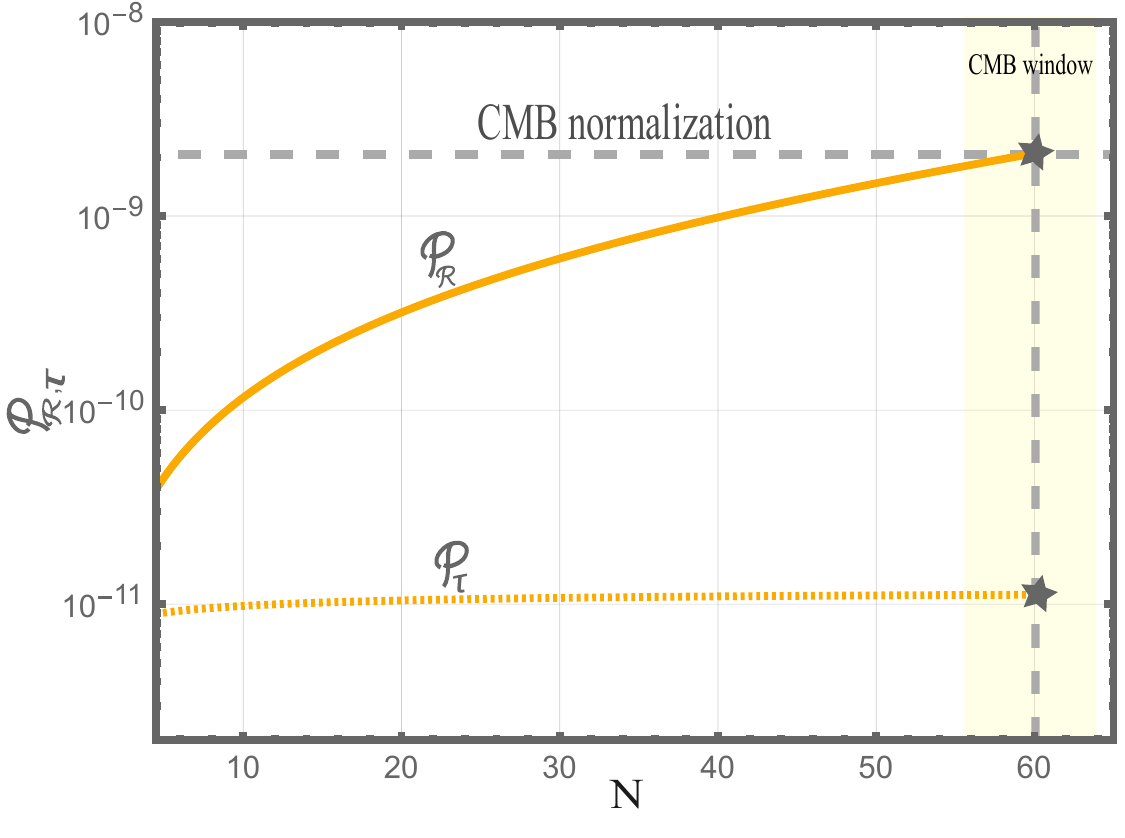}
	\caption{ The evolution of the power spectrum of curvature and tensor perturbations as a function of the number of e-folds for $V(\phi)=\phi^{2/5}$, $N= 60$ and $\beta = 0.212$ }	
	\label{figpower25}
\end{figure}
 \begin{figure}
\centering
	\includegraphics[width=0.44\textwidth]{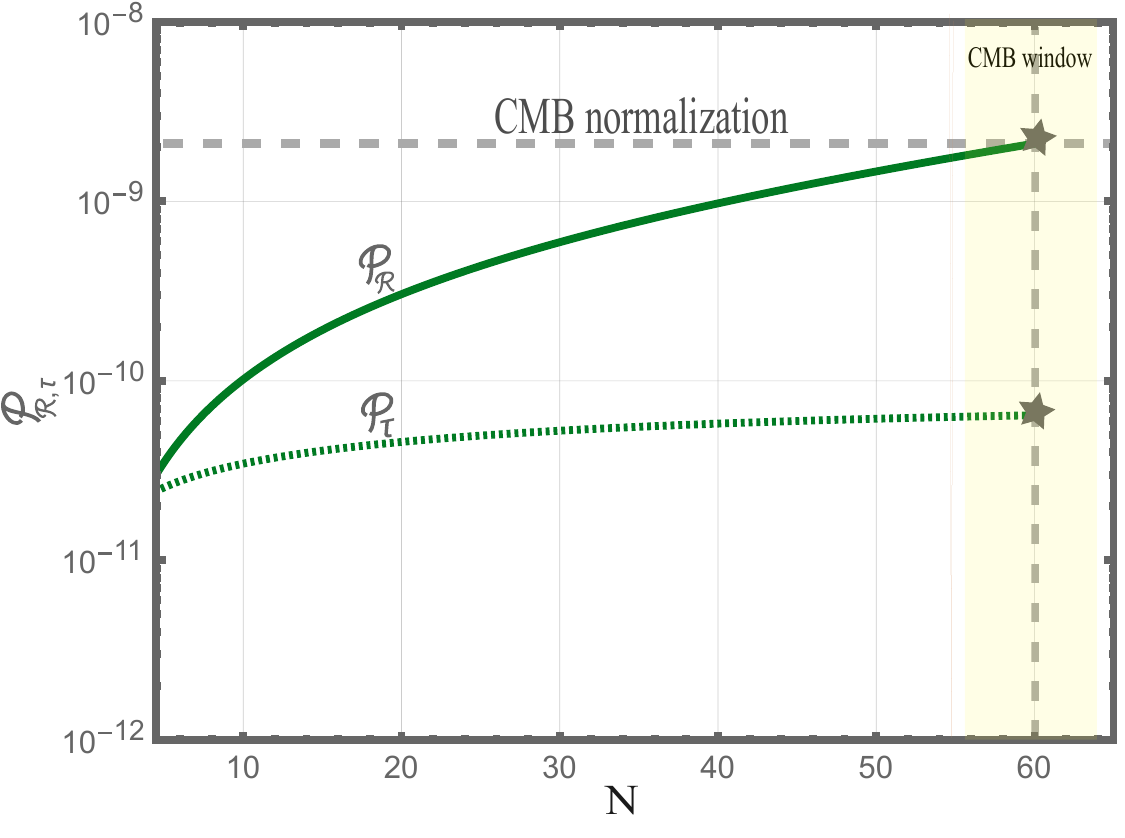}
	\caption{ The evolution of the power spectrum of curvature and tensor perturbations as a function of the number of e-folds for  $V(\phi)=\phi$, $N=60$ and $\tilde{\beta} = 0.031$.}
	\label{figpower1}
\end{figure}
Figures (\ref{figpower25}) and (\ref{figpower1}) explore the behavior of the power spectrum for curvature and tensor perturbations as a function of the number of e-folds. For this purpose, we have numerically solved Eqs. (\ref{dotH}) and (\ref{ddotphi}), using Eq. (\ref{epsilonH}),  have replaced them in Eqs. (\ref{scalarcurvsingle}) and (\ref{tensorcurvsingle}). This evolution provides crucial comprehension of the generation of primordial fluctuations and the inflation dynamics. 
In these Figures, the CMB window demonstrates the scalar field values at the moments when modes with wave numbers $k \in [0.0005,0.5]Mpc^{-1}$ cross the Hubble horizon during inflation, covering scales that the most recent CMB observations can probe. The CMB normalization ensures that theoretical inflationary models generate perturbations consistent with observational data. In the duration of inflation, the power spectrum changes as modes cross the horizon, with their amplitude being influenced by the background expansion rate and the specific properties of the single-field model. The perturbations of curvature, which seed the formation of large-scale structures, remain nearly constant on superhorizon scales.  Also, tensor perturbations are associated with primordial gravitational waves and may decay depending on the inflationary model. 
By studying the evolution of the power spectrum, we can compare theoretical predictions with observational constraints. To create a single-field inflation, like Figures (\ref{figpower25}) and (\ref{figpower1}), we need to consider $m$
and $M$ are the order in $\sim 10^{-5}$.
\\
\begin{figure}
\centering
	\includegraphics[width=0.44\textwidth]{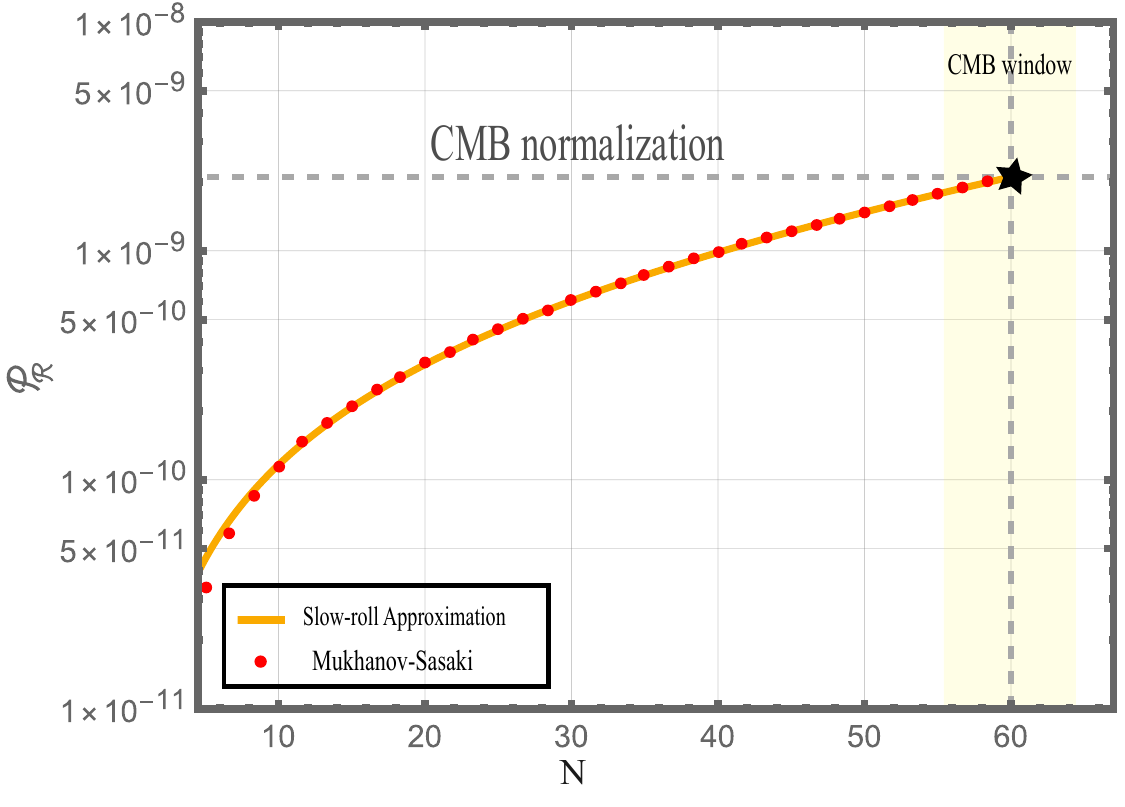}
	\caption{ The evolution of the power spectrum of curvature using the slow-roll approximation and solving the Mukhanov-Sasaki equation numerically as a function of the number of e-folds for $V(\phi)=\phi^{2/5}$, $N= 60$ and $\beta = 0.212$ }	
	\label{figsasaki25}
\end{figure}
 \begin{figure}
\centering
	\includegraphics[width=0.44\textwidth]{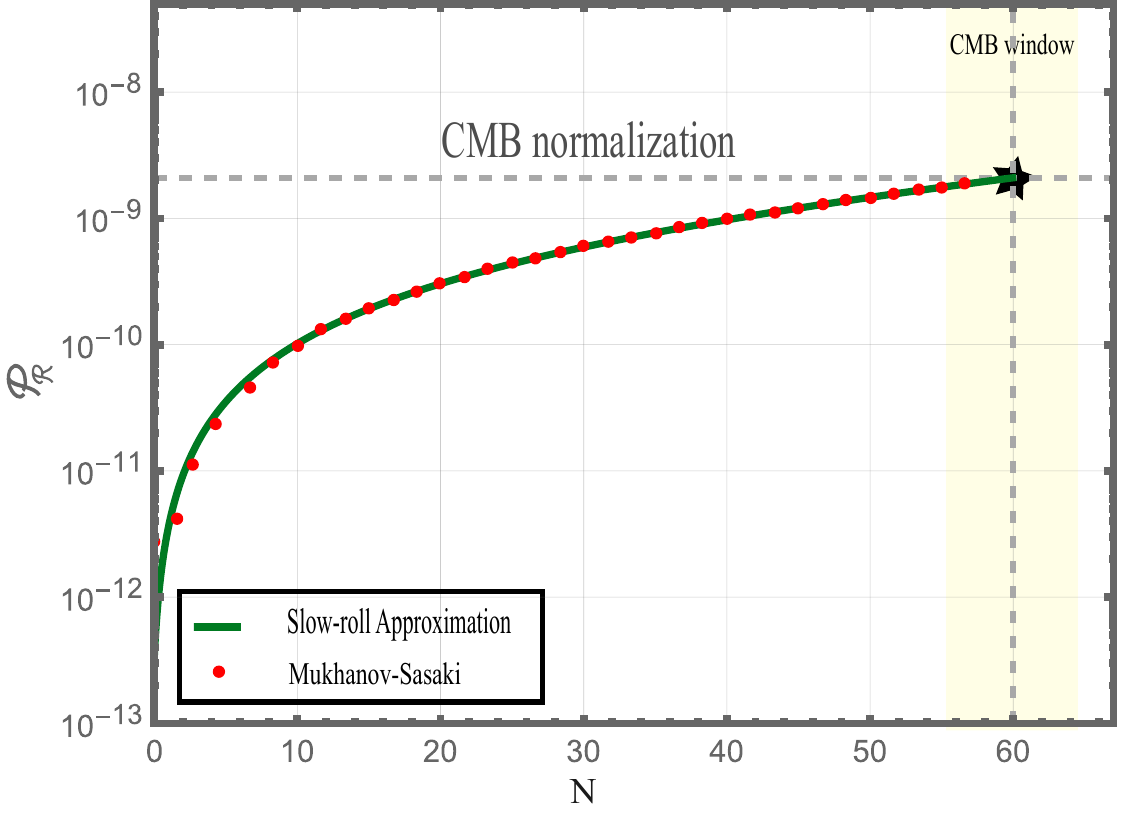}
	\caption{ The evolution of the power spectrum of curvature using the slow-roll approximation and solving the Mukhanov-Sasaki equation numerically as a function of the number of e-folds for  $V(\phi)=\phi$, $N=60$ and $\tilde{\beta} = 0.031$.}
	\label{figsasaki1}
\end{figure}
An accurate evaluation of the power spectrum of curvature perturbations can be achieved by solving the Mukhanov-Sasaki equation, which is expressed as \cite{Garriga:1999vw}:
\begin{eqnarray}\label{sasaki}
\frac{d^{2}v_{k}}{d\zeta^{2}} + \Big(c_{s}^{2}k^{2} - \frac{1}{z}\frac{d^{2}z}{d\zeta^{2}} \Big)v_{k}=0,
\end{eqnarray}
where $k$ is associated with the comoving scale through $\lambda = 2\pi /k$, and $\zeta$ denotes the conformal time. 
Moreover, the curvature perturbations $\mathcal{R}$ is associated to $v$, where  $\mathcal{R} = v/z$, and $z$ is defined as \cite{Garriga:1999vw}:
\begin{eqnarray}\label{z}
z=\frac{a^{2} \Big[\rho (X, \phi) + P(X, \phi )\Big]}{c_{s}^{2}k^{2}}.
\end{eqnarray}
Eventually, the power spectrum of the scalar mode at the horizon crossing, determined by the condition $aH =c_{s}k$, characterizes the statistical distribution of primordial curvature perturbations on different length scales. It prepares the essential understanding of the early universe's structure formation and the evolution of inflationary fluctuations, which is defined as:
\begin{eqnarray}\label{PRsasaki}
\mathcal{P}_{\mathcal{R}} = \frac{k^{3}}{2\pi} |\frac{v_{k}}{z}|^{2},
\end{eqnarray}
where $v_{k}$ is the solution of the mode function Eq. (\ref{sasaki}). Now, we have numerically solved Eqs. (\ref{dotH}) and (\ref{ddotphi}), using Eqs. (\ref{epsilonH}), (\ref{VV}), (\ref{VVV}), and have substituted them into Eq. (\ref{scalarcurvsingle}). This enabled us to numerically plot the evolution of the power spectrum of curvature using the slow-roll approximation as a function of the number of e-folds. Moreover, by solving Eq. (\ref{sasaki}) numerically and substituting into Eq. (\ref{PRsasaki}), we can plot the evolution of the power spectrum of curvature by using solving the Mukhanov-Sasaki equation numerically as a function of the number of e-folds. This evolution is illustrated in Figures. (\ref{figsasaki25}) and (\ref{figsasaki1}). In these Figures, there is a complete consistency between the power spectrum of curvature using the slow-roll approximation and solving the Mukhanov-Sasaki equation in our model.

Using the calculated power spectrum, we can calculate two important quantities of the primordial cosmological perturbations, like the tensor-to-scalar ratio, $r$, and the spectral index, $n_{s}$.  Using Eqs.  (\ref{scalarcurvsingle}) and (\ref{tensorcurvsingle}), the tensor-to-scalar ratio $r$ can obtain as \cite{Chen:2006nt}:
\begin{eqnarray}\label{rsingle1}
r \equiv \frac{\mathcal{P}_{\tau}}{\mathcal{P}_{\mathcal{R}}} =16 \varepsilon_{H}.
\end{eqnarray}
We attempt to obtain the spectral index $n_{s}$. For this purpose, we utilize the definition of $n_{s}$  \cite{Chen:2006nt}:
\begin{equation}\label{eq1}
n_{s} - 1 = \frac{d\ln \mathcal{P}_{\mathcal{R}}}{d \ln k} = \frac{d\ln \mathcal{P}_{\mathcal{R}}}{d N}. 
\end{equation}
where $k$ is the wave number and $N$ is the number of e-folds.. Using Eqs. (\ref{epsilonH}), (\ref{etaH}), and (\ref{scalarcurvsingle}) and substituting  into Eq. (\ref{eq1}), the spectral index finds \cite{Chen:2006nt}:
\begin{equation}\label{eq2}
n_{s} = 1 - 2 \varepsilon_{H} -\eta_{H}.
\end{equation}
\subsection{Analytical study of early universe parameters}
Thus far, we have derived the equations necessary for computing numerical inflationary parameters. We now intend to establish analytical equations for these parameters. 
\subsubsection{Analytical solutions of the inflationary parameters for $V(\phi) =\frac{1}{2} m^{2}\phi^{\frac{2}{5}}$}
By substituting Eq. (\ref{VV}) into Eq. (\ref{Nnew}), generally, it can be obtained the number of e-folds for the single field in our model as
\begin{eqnarray}\label{NN1}
N = \frac{15 \phi^{2}}{16} \bigg[ 1- \frac{5}{3\alpha^{4}m^{8} \phi^{\frac{8}{5}}} - \frac{5}{6\alpha^{3}m^{6} \phi^{\frac{6}{5}}}
~~~~~~~~~~~~~~ \nonumber\\
- \frac{5}{9\alpha^{2}m^{4} \phi^{\frac{4}{5}}} - \frac{5}{12\alpha m^{2} \phi^{\frac{2}{5}}} -\frac{5}{18}\alpha m^{2}\phi^{\frac{2}{5}} 
\nonumber\\
-\frac{5}{3\alpha^{5} m^{10} \phi^{2} } \log(1-\alpha m^{2}\phi^{\frac{2}{5}})
 \bigg]
\end{eqnarray}
By solving Eq. (\ref{NN1}), the scalar field $\psi$ can be obtained with respect to $N$ to the third order of $\beta =  \alpha m^{2}/2$ as  
\begin{eqnarray}\label{phi}
\phi = \frac{2N^{\frac{1}{2}}}{5^{\frac{1}{2}}} -\frac{2^{\frac{4}{5}}\times5^{\frac{1}{10}}N^{\frac{9}{10}} \beta^{2}}{7}  - \frac{N^{\frac{11}{10}}\beta^{3}
}{2^{\frac{4}{5}}\times 5^{\frac{1}{10}}} + \mathcal{O}(\beta^{4}).~~
\end{eqnarray} 
By substituting Eqs. (\ref{f}), (\ref{epsilonH}), (\ref{VV}), and (\ref{phi}) into Eq. (\ref{rsingle1}), we can acquire the relation between the tensor-to-scalar ratio and the number of e-folds $N$ to the third order of $\beta$ as
\begin{eqnarray}\label{rA}
r \simeq \frac{8}{5N} -\frac{8 \times 2^{\frac{2}{5}} \beta}{5 \times 5^{\frac{1}{5}}N^{\frac{4}{5}}} - \frac{72\times 2^{\frac{4}{5}} \beta^{2}}{ 35 \times 5^{\frac{2}{5}}N^{\frac{3}{5}}}
- \frac{148 \times 2^{\frac{1}{5}} \beta^{3}}{ 35 \times 5^{\frac{3}{5}}N^{\frac{2}{5}}} 
\nonumber\\
+ \mathcal{O}(\beta^{4}).~~~~~~~~~~~~~~~~~~~~~~~~~~~~~~~~~~~~~~~~~~~~~~~~~~
\end{eqnarray} 
By substituting Eqs. (\ref{f}), (\ref{epsilonH}), (\ref{etaH}), (\ref{VV}), and (\ref{phi}) into Eq. (\ref{eq2}), we can acquire the relation between the spectral index and the number of e-folds, $N$, to the third order of $\beta$ as
\begin{eqnarray}\label{nsA}
n_{s} \simeq  1 - \frac{6}{5N} -\frac{16\times 2^{\frac{4}{5}} \beta^{2}}{35 \times 5^{\frac{2}{5}}N^{\frac{3}{5}}} - \frac{3\times 2^{\frac{1}{5}} \beta^{3}}{ 5^{\frac{3}{5}}N^{\frac{2}{5}}}
+ \mathcal{O}(\beta^{4}).
\end{eqnarray} 
\subsubsection{Analytical solutions of the inflationary parameters for $V(\phi) =\frac{1}{2} M^{2}\phi$}
By substituting Eq. (\ref{VVV}) into Eq. (\ref{Nnew}), generally, it can be obtained the number of e-folds for the single field in our model as
\begin{eqnarray}\label{NN2}
N = \frac{3 \phi^{2}}{8} \bigg[ 1- \frac{2}{3\alpha M^{2} \phi} -\frac{2}{9}\alpha M^{2}\phi 
~~~~~~~~~~~~~~~~~~~~\nonumber\\
-\frac{2}{3\alpha^{2}M^{4} \phi^{2} } \log(1-\alpha M^{2}\phi).
 \bigg]
\end{eqnarray}
By solving Eq. (\ref{NN2}), the scalar field $\psi$ can be obtained with respect to $N$ to the third order of $\tilde{\beta} =  \alpha M^{2}/2$ as  
\begin{eqnarray}\label{phi1}
\phi =( 2 N)^{\frac{1}{2}} - \frac{N^{\frac{3}{2}}\tilde{\beta}^{2}}{ 2^{\frac{1}{2}}} -\frac{8 N^{2}\tilde{\beta}^{3}}{5} + \mathcal{O}(\tilde{\beta}^{4}).
\end{eqnarray} 
By substituting Eqs. (\ref{f}), (\ref{epsilonH}), (\ref{VVV}), and (\ref{phi1}) into Eq. (\ref{rsingle1}), we can acquire the relation between the tensor-to-scalar ratio and the number of e-folds $N$ to the third order of $\tilde{\beta}$ as
\begin{eqnarray}\label{rA1}
r \simeq \frac{4}{N} -\frac{4 \times 2^{\frac{1}{2}} \tilde{\beta}}{N^{\frac{1}{2}}} - 12\tilde{\beta}^{2} -\frac{58\times 2^{\frac{1}{2}}N^{\frac{1}{2}}\tilde{\beta}^{3}}{5} + \mathcal{O}(\tilde{\beta}^{4}).~~~
\end{eqnarray} 
By substituting Eqs. (\ref{f}), (\ref{epsilonH}), (\ref{etaH}), (\ref{VVV}), and (\ref{phi1}) into Eq. (\ref{eq2}), we can acquire the relation between the spectral index and the number of e-folds, $N$, to the third order of $\beta$ as
\begin{eqnarray}\label{nsA1}
n_{s} \simeq  1 - \frac{3}{2N} - \frac{5 \tilde{\beta}^{2}}{2} - \frac{42\times 2^{\frac{1}{2}} N^{\frac{1}{2}}\tilde{\beta}^{3}}{ 5} + \mathcal{O}(\tilde{\beta}^{4}).
\end{eqnarray} 

\begin{table}[h!]
\centering
\scriptsize
{\tiny
\setlength{\arrayrulewidth}{0.3mm}
\begin{tabular}{|c| c| c| c|} 
 \hline
Model. &            $N$&                     Range                                                 & Observational constraints \\ 
 \hline
           &     50          &         $0.159477  \leq  \beta <  0.214459 $             &  $ 68\%$~Planck~ TT, TE, EE + low E + \\ 
           &                   &                                                                                &   lensing + BK18 + BAO.  \\ 
           &                   &                                                                                &    \\
$\phi^{2/5}$ &  60    &         $0.192211   <\beta < 0.226399$                    &  $ 68\%$~Planck~ TT, TE, EE + low E + \\
           &                   &                                                                                &  lensing + BK18 + BAO. \\   
           &                   &                                                                                &    \\

           &    70           &         $ 0.212045 \leq \beta < 0.23141 $                 &  $ 68\%$~Planck~ TT, TE, EE + low E +  \\
           &                   &                                                                                &  lensing + BK18 + BAO. \\     
           &                   &                                                                                &    \\               
  \hline
          &      50         &         $0.00894551  <\tilde{\beta}<0.0343115 $       & $68\%$~Planck~ TT, TE, EE + low E + \\
          &                   &                                                                                 &  lensing. ~~~~~~~~~~~~~~~~~~~~~~ \\  
          &                   &                                                                                 &    \\
$\phi$&       60        & $ 0.026033 <\tilde{\beta}< 0.0343436$                   & $ 95\%$~Planck~ TT, TE, EE + low E +  \\ 
          &                   &                                                                                 & lensing + BK18 + BAO.  \\ 
          &                   &                                                                                 &    \\
          &       70        & $ 0.031338 \leq \tilde{\beta}< 0.0343121$              & $ 68\%$~Planck~ TT, TE, EE + low E +  \\ 
          &                   &                                                                                 &  lensing + BK18 + BAO.  \\ 
  \hline
\end{tabular}
}
\caption{The $\beta$ and $\tilde{\beta}$ ranges in the single-field model by considering the potentials  $\phi^{2/5}$ and $\phi$.}
\label{tab1}
\end{table}

\begin{figure}
\centering
	\includegraphics[width=0.46\textwidth]{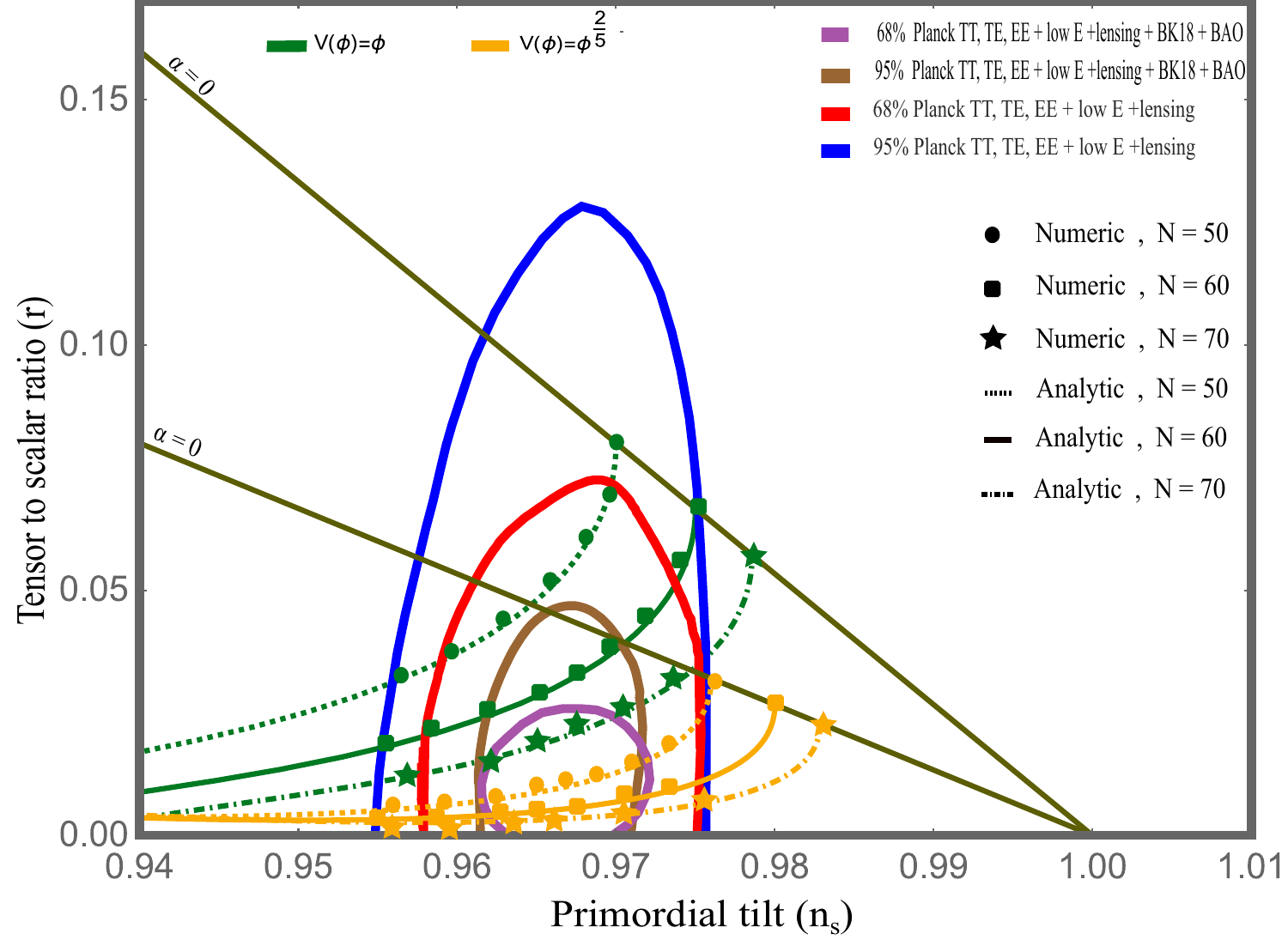}
	\caption{ The variation of the tensor-to-scalar ratio with respect to the spectral index is shown both numerically and analytically for the different values of $\beta$ and $\tilde{\beta}$. Each colored curve corresponds to the single-field model $\alpha \neq 0$ by considering the potentials  $\phi^{2/5}$ and $\phi$ for the number of e-folds $50, 60, 70$. The solid black line represents the standard canonical single-field inflationary model $(\alpha=0)$ for the number of e-folds $ 50, 60, 70$.}
	\label{fig1}
\end{figure}
\section{Numerical solutions vs. analytical predictions}\label{sec4}
In this section, we aim to compare the numerical solutions with the analytical predictions for the key inflationary parameters in our model with the potentials considered in Eqs. (\ref{VV}) and (\ref{VVV}). We investigate how these results align with observational constraints such as $Planck~ TT, TE, EE + low E + lensing + BK18 + BAO$ and $Planck~ TT, TE, EE + low E + lensing$ with $68\%$ and $95\%$ confidences \cite{BICEP:2021xfz}. This comparison prepares a valid understanding of theoretical approximations and numerical solutions to describe inflationary dynamics for our model. Also, we figure out the differences between our model and the standard canonical single-field model.

Recently, the findings from $Planck$ remarkably restrict the limitation of acceptable inflationary parameters in the early era. For instance, the observational data from $Planck$ and $BICEP/Keck$ point out that there is a bound  $r<0.036$ with $95\%$ confidence for the tensor-to-scalar ratio parameter \cite{BICEP:2021xfz}. The corrections from $\beta$ and $\tilde{\beta}$ improve our model's tensor-to-scalar ratio and spectral index values compared to the standard canonical single-field model. These corrections lead to a reduction in the values of $r$ and $n_s$. For example, to remain consistent with large-scale CMB observations from the Planck TT, TE, EE + lowE + lensing + BK15 + BAO data at the pivot scale $k_{*}$ \cite{Planck:2018vyg}:
\begin{equation}
0.956<n_{s}<0.978 ~~~~,~~~~r(k_{*})\leq 0.066
\end{equation}
Using Eqs. (\ref{rA}), (\ref{nsA}), (\ref{rA1}), and (\ref{nsA1}), we can obtain the $\beta$ and $\tilde{\beta}$ ranges for our model along with observational constraints from $Planck$ and $BICEP/Keck$ which is shown in Table (\ref{tab1}) for the number of e-folds $N = 50, 60, 70$. As shown in this Table, the presence of $\beta$ and $\tilde{\beta}$ influences the tensor-to-scalar ratio and spectral index values. These findings indicate that our model is consistent with observational constraints from  $Planck$ and $BICEP/Keck$ with $68\%$ and $95\%$ confidences.
\begin{figure}
\centering
	\includegraphics[width=0.44\textwidth]{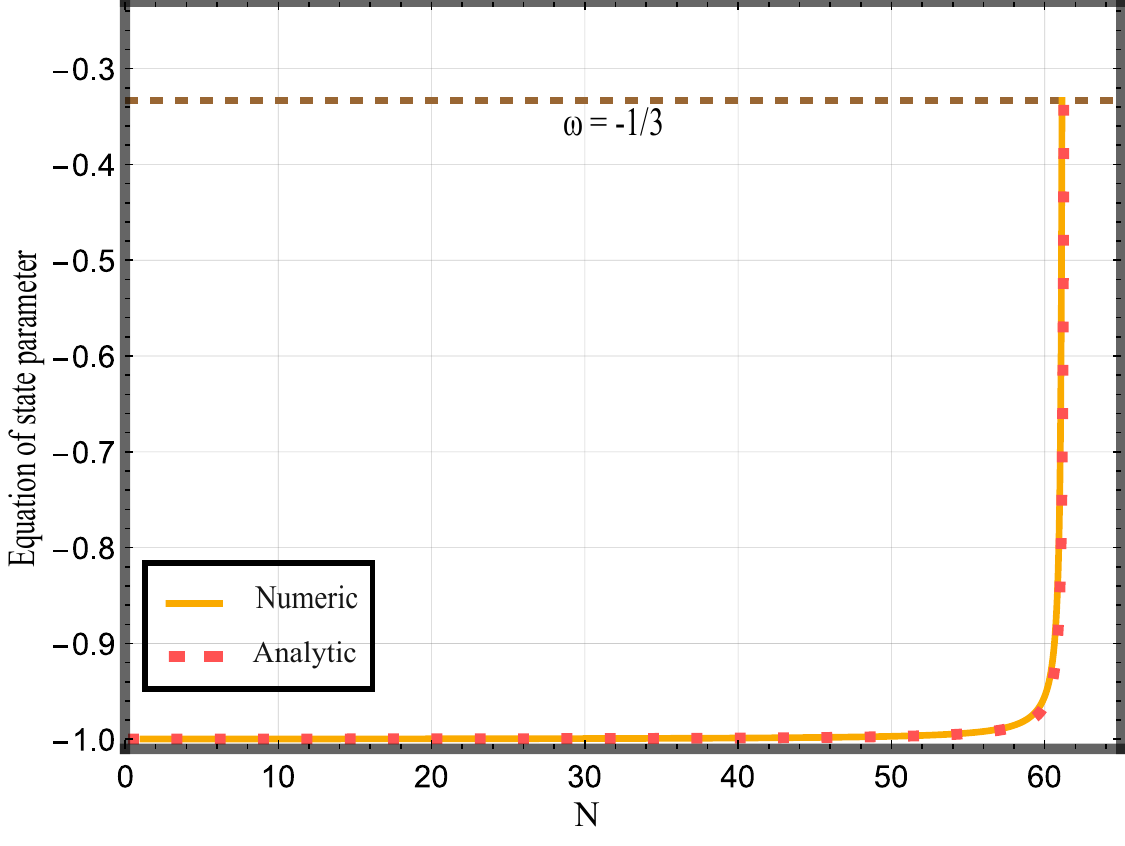}
	\caption{ The evolution of the equation of state parameter as a function of the number of e-folds is shown both numerically and analytically for $V(\phi)=\phi^{2/5}$, $N= 60$ and $\beta = 0.212$ }	
	\label{fig2}
\end{figure}
 \begin{figure}
\centering
	\includegraphics[width=0.44\textwidth]{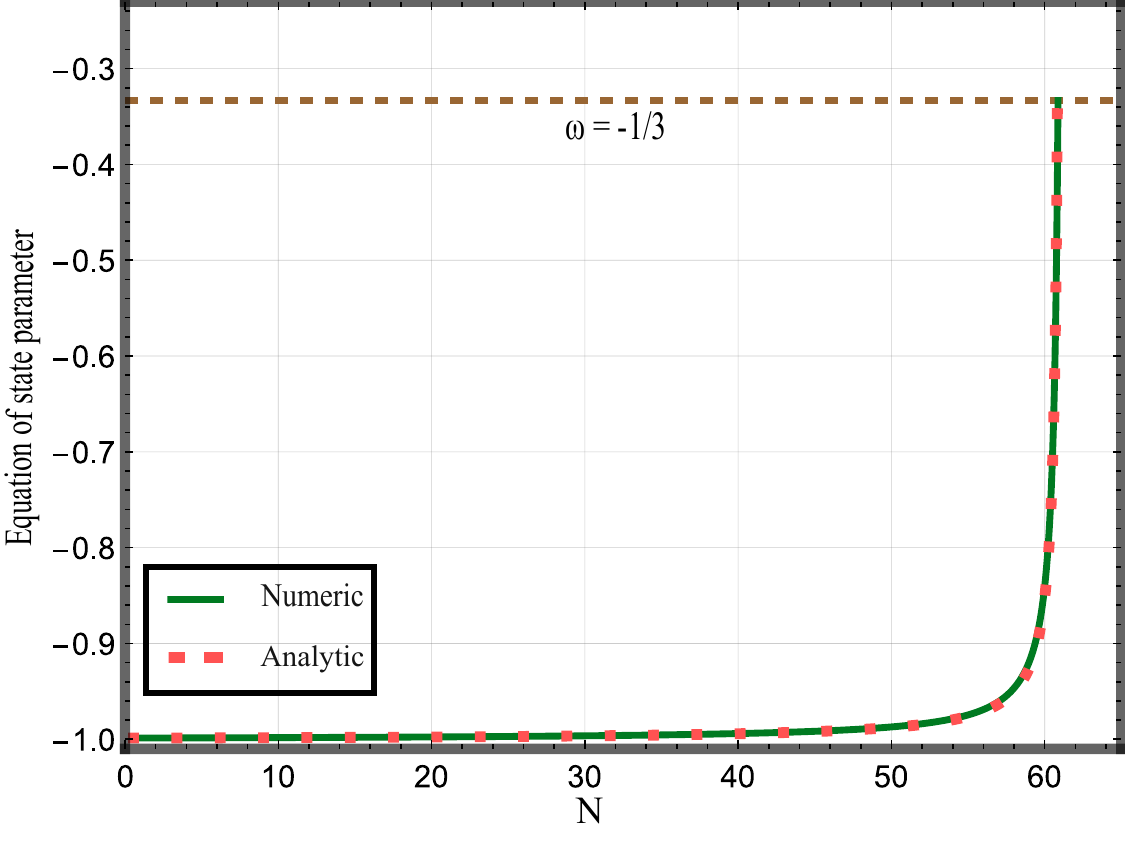}
	\caption{ The evolution of the equation of state parameter as a function of the number of e-folds is shown both numerically and analytically for $V(\phi)=\phi$, $N=60$ and $\tilde{\beta} = 0.031$.}
	\label{fig3}
\end{figure}
\begin{figure}
\centering
	\includegraphics[width=0.44\textwidth]{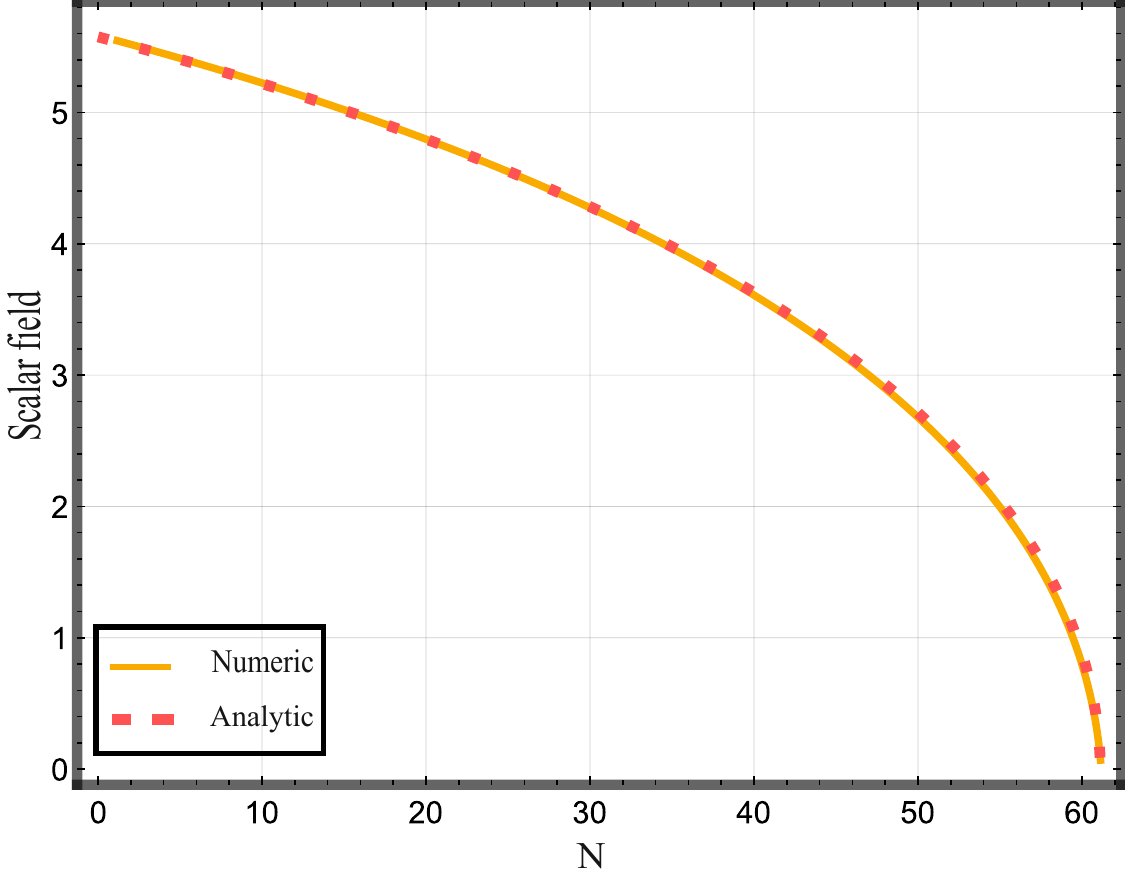}
	\caption{ The evolution of the scalar field as a function of the number of e-folds is shown both numerically and analytically for $V(\phi)=\phi^{2/5}$, $N= 60$ and $\beta = 0.212$ }	
	\label{figscalarfield25}
\end{figure}
 \begin{figure}
\centering
	\includegraphics[width=0.44\textwidth]{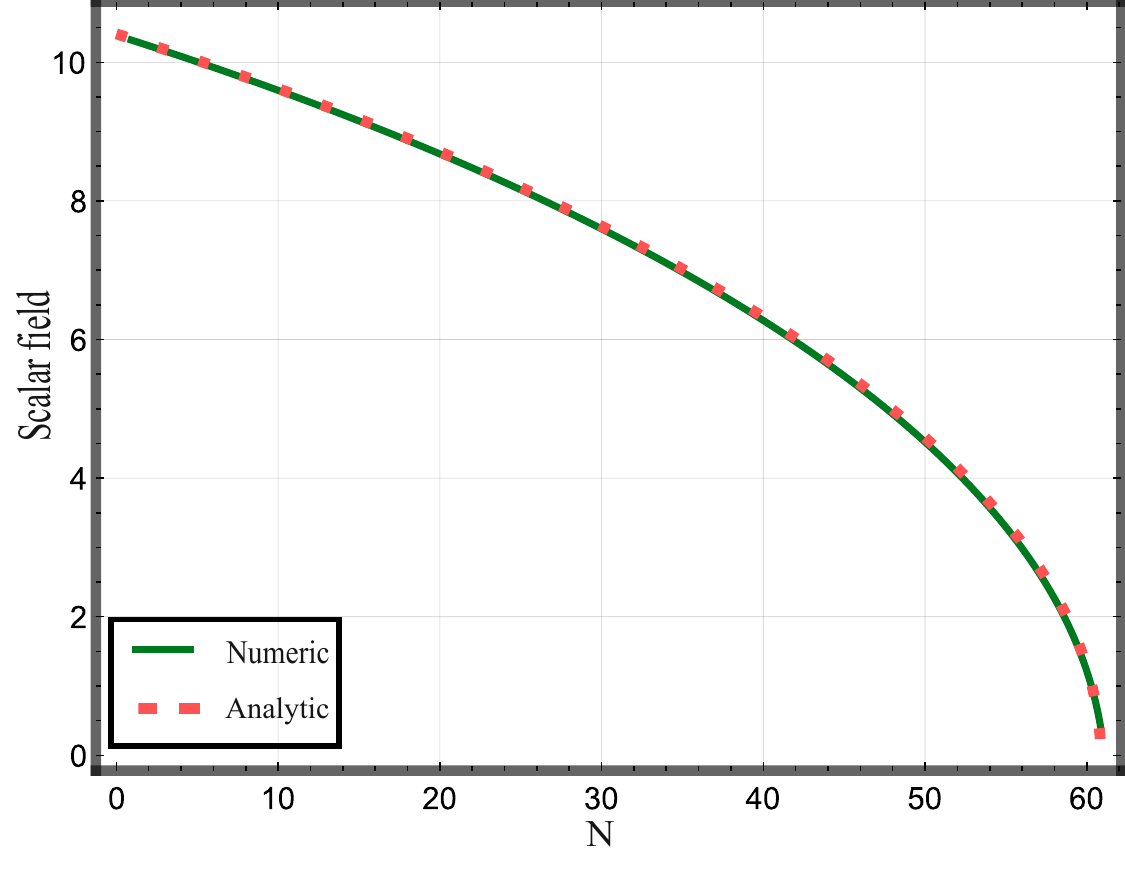}
	\caption{ The evolution of the scalar field as a function of the number of e-folds is shown both numerically and analytically for $V(\phi)=\phi$, $N=60$ and $\tilde{\beta} = 0.031$.}
	\label{figscalarfield1}
\end{figure}
\begin{figure}
\centering
	\includegraphics[width=0.46\textwidth]{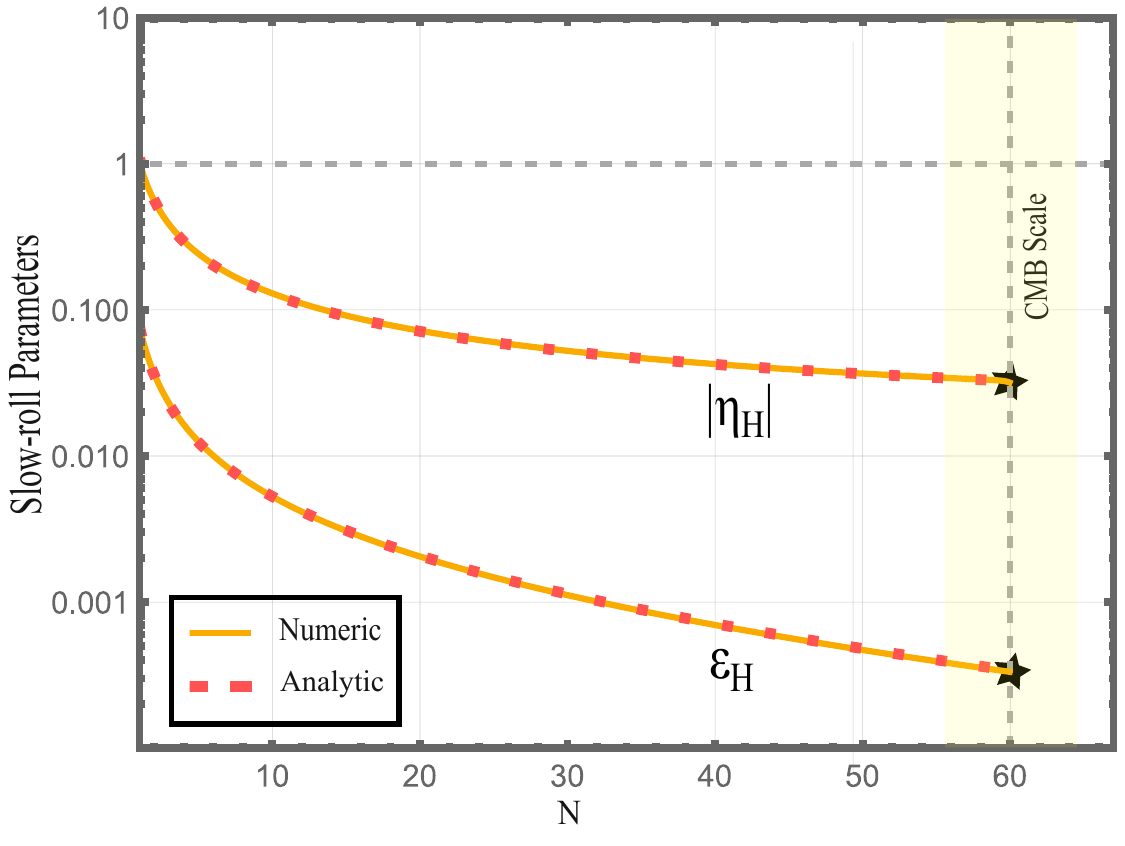}
	\caption{ The evolution of the slow-roll parameters as a function of the number of e-folds is shown both numerically and analytically for $V(\phi)=\phi^{2/5}$, $N= 60$ and $\beta = 0.212$ }	
	\label{figSR25}
\end{figure}
 \begin{figure}
\centering
	\includegraphics[width=0.46\textwidth]{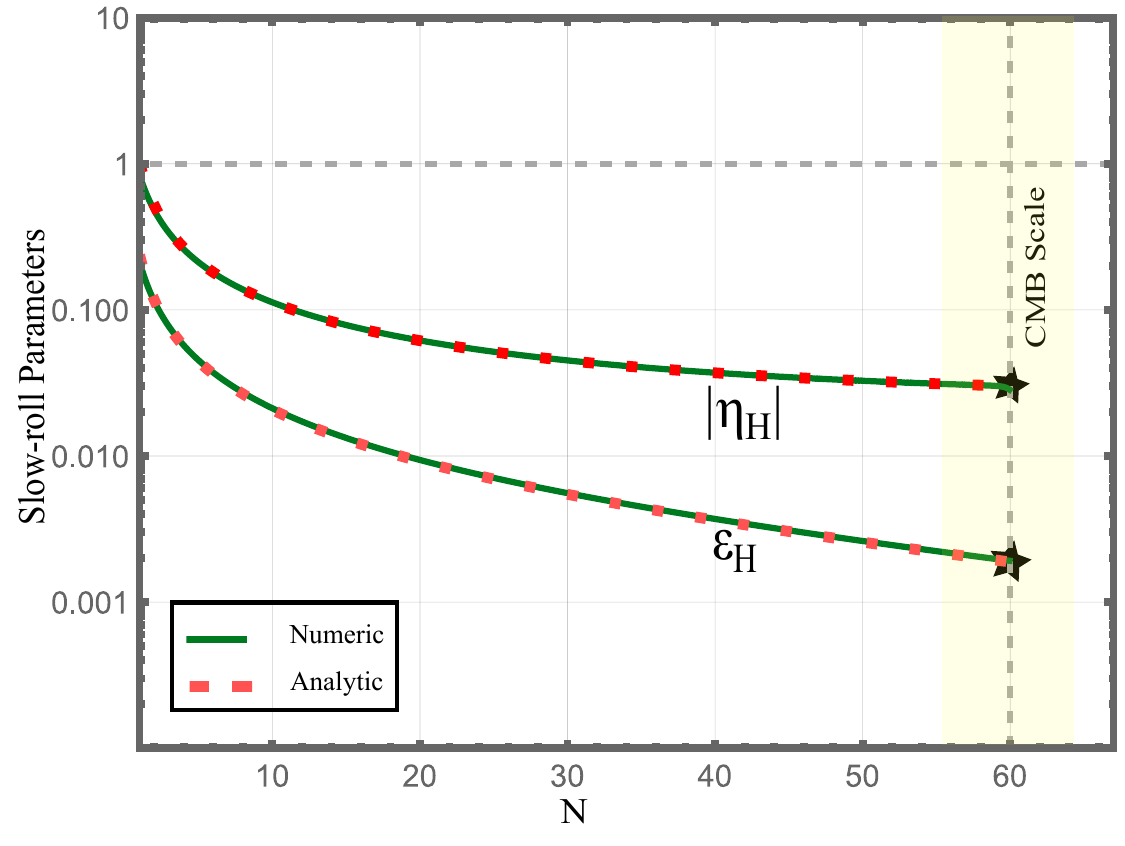}
	\caption{ The evolution of the slow-roll parameters as a function of the number of e-folds is shown both numerically and analytically for $V(\phi)=\phi$, $N=60$ and $\tilde{\beta} = 0.031$.}
	\label{figSR1}
\end{figure}

To present the results in Table I, we have numerically solved Eqs. (\ref{dotH}) and (\ref{ddotphi}), using Eqs. (\ref{epsilonH}) and (\ref{etaH}), and have replaced them in Eqs. (\ref{rsingle1}) and (\ref{eq2}). This allowed us to numerically plot the tensor-to-scalar ratio as a function of the spectral index in Figure. (\ref{fig1}). Furthermore, by using the results from this table and substituting them into Eqs. (\ref{rA}), (\ref{nsA}), (\ref{rA1}), and (\ref{nsA1}), we have analytically illustrated the tensor-to-scalar ratio as a function of the spectral index in this Figure. As observed, our model indicates that the presence of $\beta$ and $\tilde{\beta}$ affects the values of the tensor-to-scalar ratio and spectral index and is consistent with observational constraints from  $Planck$ and $BICEP/Keck$ with $68\%$ and $95\%$ confidences. In contrast, the results of the standard canonical single-field model with the potentials $\phi^{\frac{2}{5}}$ for $N=50, 60, 70$ and $\phi$ for $N=60, 70$ were ruled out and are inconsistent with observational constraints from  $Planck$ and $BICEP/Keck$.
It is important to note that we expand up to the fourth order in this work; however, to ensure a strong agreement between numerical and analytical results, we also consider higher-order terms, such as the tenth order.

We have numerically solved Eqs. (\ref{dotH}) and (\ref{ddotphi}), using Eqs. (\ref{epsilonH}), (\ref{VV}), (\ref{VVV}), and have substituted them into Eq. (\ref{omega1}). This enabled us to numerically plot the equation of state parameter as a function of the number of e-folds in Figures. (\ref{fig2}) and (\ref{fig3}). Furthermore, using Eqs. (\ref{epsilonH}), (\ref{VV}), (\ref{VVV}) and substituting them into Eq. (\ref{omega1}), we have analytically illustrated the equation of state parameter as a function of the number of e-folds in these Figures. 
As shown, the equation of state parameter demonstrates the relation between the energy density and the pressure of the universe in the inflationary phase. It generally remains almost $-1$, which mentions that a form of energy dominates the cosmic with approximately zero pressure and leads to the rapid expansion of inflation. After finishing the inflation, $\epsilon_{H} =1$, this parameter approaches $-1/3$, which confirms this result from Eq. (\ref{omega1}) as well. Additionally, the analytical and numerical results of the equation of state parameter illustrate remarkable consistency for our model.

As mentioned, the inflaton field is responsible for driving the accelerated expansion in an inflationary universe. By solving Eqs. (\ref{dotH}) and (\ref{ddotphi}), using Eqs. (\ref{phi}) and (\ref{phi1}), we can plot the evolution of the scalar field as a function of the number of e-folds for our model, which has been shown in Figures (\ref{figscalarfield25}) and (\ref{figscalarfield1}). In these Figures, there is a complete consistency between the analytical and numerical results of the scalar field in our model.

The slow-roll parameters are necessary for qualifying the evolution of the early universe and warranting that it goes forward under controlled conditions. These parameters specify how the scalar field rolls down to its minimum potential smoothly. The slow roll parameter, $\varepsilon_{H}$, measures the changes of the Hubble parameter in the duration of inflation and indicates when the inflation will finally end, $\varepsilon_{H}=1$. The slow-roll parameter $\eta_{H}$ controls the curvature of the potential. During inflation, these parameters must remain small to keep a period of accelerated expansion, $\varepsilon_{H}, \eta_{H} \ll 1$. It is important to mention that these parameters directly impact observable quantities, including the spectral index and the tensor-to-scalar ratio. By solving Eqs. (\ref{dotH}) and (\ref{ddotphi}), using Eqs. (\ref{f}),  (\ref{VV}), (\ref{VVV}), (\ref{phi}) and (\ref{phi1}), and substituting into Eq. (\ref{epsilonH}) and (\ref{etaH}), we can plot the evolution of the slow-roll parameters as a function of the number of e-folds for our model, which has been shown in Figures (\ref{figSR25}) and (\ref{figSR1}). In these Figures, there is a complete consistency between the analytical and numerical results of the scalar field in our model.
\section{Conclusion}\label{sec5}
In this work, we have studied a specific class of the K-essence models incorporating a coupling term, $\alpha$, between canonical Lagrangian and the potential. This coupling term could change the slow-roll dynamics and have impacts on key inflationary parameters. The initial purpose of this study has been to analyze the effect of the $\alpha$ term on the parameters, including the Hubble parameter, slow-roll parameters, equation of state parameter, power spectrum, tensor-to-scalar ratio, and spectral index. Our findings have shown that the $\alpha$ term could influence these parameters. The presence of $\alpha$ is important as this term has made our model better consistent with observational constraints from ${\rm Planck~ TT, TE, EE + low E + lensing + BK18 + BAO}$ and ${\rm Planck~ TT, TE, EE + low E + lensing}$ with $68\%$ and $95\%$ confidences. 

By solving the equations numerically and deriving analytical results, we have compared the numerical solutions with the analytical predictions for the key inflationary parameters in our model with the potentials $\phi$ and $\phi^{2/5}$. This comparison has prepared a valid understanding of theoretical approximations and numerical solutions to describe inflationary dynamics for the studied model. We have introduced the the $\beta$ and $\tilde{\beta}$ which are relevant to the $m$, $M$, $\beta$ and $\tilde{\beta}$. In addition, we have obtained the best ranges of $\beta$ and $\tilde{\beta}$ along with observational constraints from ${\rm Planck}$ and ${\rm BICEP/Keck}$, which is shown in Table (\ref{tab1}) for the number of e-folds $N = 50, 60, 70$. 

Finally, we have obtained the differences between our model and the standard canonical single-field model, which is shown in Figure (\ref{fig1}). The standard canonical single-field model, for the considered potential in this work, is ruled out with observational constraints, but our model is completely consistent with observational constraints from ${\rm Planck}$ and ${\rm BICEP/Keck}$.
\section*{Acknowledgments}
This paper is supported by the NSRF via the Program Management Unit for Human Resources $\&$ Institutional Development, Research and Innovation [grant number B13F670063].

\bibliography{single-field-inflation}

\end{document}